Stanislav Dolgopolov
dolgopolov-s@list.ru

# Formation of Cooper Pairs as a Consequence of Exchange Interaction

**Abstract**: Analyzing the exchange energy of two conduction electrons in a crystal we find that the exchange energy may be negative and, thus, a singlet state may be favorable. A full overlap in real space of wave functions of two conduction electrons leads to a deeper exchange energy. Thus, the exchange interaction causes a bond between two conduction electrons in real space. The singlet bond is possible because the singlet electrons can simultaneously and permanently occupy one spatial ground state, so the average energy of paired electrons is lower than the energy of unpaired electrons. Thus, the pairing is a result of the Pauli Exclusion Principle. If conduction electrons, before pairing, are put on the Fermi surface in the momentum space, then every pair may exist permanently in time. The motion of conduction electrons in the crystal may prevent the formation of Cooper pairs, because the kinetic energy of the motion is usually larger than the binding energy in the pair. Conduction electrons as standing waves are local and have zero momenta, hence their momenta are synchronous; therefore, weak singlet bonds are stable despite the large kinetic energy on the Fermi surface. The local pairing of standing waves explains the inverse isotope effect, $T_c$ – dome, insulator-superconductor transitions and many other facts about superconductors. The electron pairs, as bosons, can form a macroscopically coherent Bose-Einstein-Condensate and, thus, become non-dissipative and non-local.

## 1. Introduction and motivation.

The knowledge of root causes of superconductivity (SC) would explain many mysterious facts about all classes of superconductors. However, a unified solution remains still an open question, current theories are not universal and explain many effects ambiguously [1]. The mainstream theories assume that SC is a result of the electron pairing at a mean-field approximation, the spin ordering plays a part for the pair formation [2], [3], [4]. Well known is that every spin ordering is related with the exchange interaction, which influences the total energy of the electrons interacting with every particle of the crystal. Moreover the exchange interaction may in itself cause binding states in quantum systems [5], [6], [7]. Therefore, the role of the exchange interaction is crucially important for the pair formation. Considering electrons as standing waves in the field of the crystal, we can show that the exchange interaction may in principle lead to binding states of conduction electrons. Under certain conditions the paired electrons become superconducting.

## 2. Formation of superconducting pairs.

Normally the spins of conduction electrons in a crystal are unordered because the thermal fluctuations and own motion of electrons destroy the spin ordering. Thus, the spin of every conduction electron $e_1$ is random to spin of each other electron. This state of electron $e_1$ is designated as **unpaired** or **single**. If the spins of electrons $e_1$, $e_2$ form a stable **singlet** in their overlap area in real space, then the state of the electrons is designated as **paired**.

Each conduction electron at each moment of time occupies an eigenstate in the crystal. Knowing the eigenstates of unpaired electrons $e_1$, $e_2$ we can compute their exchange energy.

If electrons $e_1$, $e_2$ form a singlet, then their overall position-space wave function $\Phi(\vec{r}_1,\vec{r}_2)$ is symmetric:

$$\Phi(\vec{r}_1,\vec{r}_2)=\frac{1}{\sqrt{2}}\Psi_1(\vec{r}_1)\Psi_2(\vec{r}_2)+\frac{1}{\sqrt{2}}\Psi_1(\vec{r}_2)\Psi_2(\vec{r}_1) \qquad (2.1)$$

where $\Psi_1(\vec{r}_1)$, $\Psi_2(\vec{r}_2)$ are eigenstates of unpaired $e_1$, $e_2$ just before pairing; $\vec{r}_1,\vec{r}_2$ are radius-vectors of $e_1$, $e_2$.

The sum of the direct and exchange energies ($D+J$ ) we find substituting $\Phi(\vec{r}_1,\vec{r}_2)$ from Eq. (2.1) into the integral with an overall energy operator $\hat{O}(\vec{r}_1,\vec{r}_2)$:

$$D+J=\left\langle \Phi(\vec{r}_1,\vec{r}_2)\middle|\hat{O}(\vec{r}_1,\vec{r}_2)\middle|\Phi(\vec{r}_1,\vec{r}_2)\right\rangle=$$

$$=\left\langle \Psi_1(\vec{r}_1)\Psi_2(\vec{r}_2)\middle|\hat{O}(\vec{r}_1,\vec{r}_2)\middle|\Psi_1(\vec{r}_1)\Psi_2(\vec{r}_2)\right\rangle+\left\langle \Psi_1(\vec{r}_1)\Psi_2(\vec{r}_2)\middle|\hat{O}(\vec{r}_1,\vec{r}_2)\middle|\Psi_1(\vec{r}_2)\Psi_2(\vec{r}_1)\right\rangle \qquad (2.2)$$

If the exchange interaction of electrons $e_1$, $e_2$ is very weak, then their spins remain unpaired and fully random. In this case the exchange energy is negligible and the overall energy of the unpaired electrons is equal to the direct term $D$, as it should be for the unpaired electrons in eigenstates $\Psi_1(\vec{r}_1)$, $\Psi_2(\vec{r}_2)$:

$$\left\langle E(\vec{r}_1,\vec{r}_2)\right\rangle=D=\left\langle \Psi_1(\vec{r}_1)\Psi_2(\vec{r}_2)\middle|\hat{O}(\vec{r}_1,\vec{r}_2)\middle|\Psi_1(\vec{r}_1)\Psi_2(\vec{r}_2)\right\rangle \qquad (2.3)$$

The exchange energy $J$ from Eq. (2.2) is

$$J = \left\langle \Psi_1(\vec{r}_1)\Psi_2(\vec{r}_2) \middle| \hat{O}(\vec{r}_1,\vec{r}_2) \middle| \Psi_1(\vec{r}_2)\Psi_2(\vec{r}_1) \right\rangle \qquad (2.4)$$

If the exchange interaction of electrons $e_1$, $e_2$ is not negligible, then their exchange term $J$ is not zero. A singlet state of $e_1$, $e_2$ is favorable, if their exchange energy $J$ is negative [8]. $J$ takes into account the modification of the initially unpaired wave functions resulting from the pairing. This wave function modification influences all interactions of $e_1$, $e_2$ in the crystal; hence we must compute the exchange energy $J$ for $\hat{O}(\vec{r}_1,\vec{r}_2)$ as a sum of all interactions of $e_1$, $e_2$ including their kinetic energy, repulsion of $e_1$, $e_2$ from every conduction electron and attraction of $e_1$, $e_2$ to every ion.

If two electrons form a triplet, then their overall position-space wave function is antisymmetric. The triplet state of $e_1$, $e_2$ is favorable, if their exchange energy $J$ is positive.

We assume that $\Psi_1(\vec{r}_1)$, $\Psi_2(\vec{r}_2)$ have an overlap in real space and contain similar atom orbitals (for sample s-orbitals); momenta of $e_1$, $e_2$ along the crystal are equal or zero. In this case $\Psi_1(\vec{r}\,)$, $\Psi_2(\vec{r}\,)$ are **not orthogonal** as orbitals of the ground state in $H_2$-molecule or in Helium atom; hence the overlap integral $\left\langle \Psi_1(\vec{r}) \middle| \Psi_2(\vec{r}) \right\rangle$ appearing in Eq. (2.4) is not zero:

$$\left\langle \Psi_1(\vec{r}) \middle| \Psi_2(\vec{r}) \right\rangle \neq 0 \qquad (2.5)$$

Wave functions of conduction electrons fade out slowly with the distance and can cover many ions. Thus, the wave functions of electrons may overlap in a shared real space, so we consider at first the limiting case that $\Psi_1(\vec{r}\,)$, $\Psi_2(\vec{r}\,)$ almost coincide in real space, i.e.:

$$\Psi_1(\vec{r}\,) \approx \Psi_2(\vec{r}\,) \qquad (2.6)$$

Below we will see that this assumption is true because a maximal overlap in real space of two paired wave functions is energetically favorable in comparison to a partial overlap.

Using Eq. (2.6) and non-orthogonality of $\Psi_1(\vec{r}\,)$, $\Psi_2(\vec{r}\,)$ in Eq. (2.5) we can use for Eq. (2.4):

$$\Psi_1(\vec{r}_2) = \Psi_2(\vec{r}_2), \quad \Psi_2(\vec{r}_1) = \Psi_1(\vec{r}_1) \qquad (2.7)$$

Substituting Eq. (2.7) into (2.4) we obtain:

$$J = \left\langle \Psi_1(\vec{r}_1)\Psi_2(\vec{r}_2) \middle| \hat{O}(\vec{r}_1,\vec{r}_2) \middle| \Psi_1(\vec{r}_1)\Psi_2(\vec{r}_2) \right\rangle \qquad (2.8)$$

We see that Eq. (2.8) is equal to Eq. (2.3), i.e. in case of the full overlap of non-orthogonal wave functions $\Psi_1(\vec{r}\,)$, $\Psi_2(\vec{r}\,)$ the exchange energy of two singlet electrons is equal to the overall energy of two initially unpaired electrons:

$$J = \left\langle E(\vec{r}_1,\vec{r}_2) \right\rangle \qquad (2.9)$$

We may define that the electron energy outside of the crystal is zero. Then the electron energy $\left\langle E(\vec{r}_1,\vec{r}_2) \right\rangle$ inside the crystal should be negative, otherwise states of single electrons in the crystal are instable. Thus, the exchange energy $J$ of $e_1$, $e_2$ is also negative and the paired state is favorable in comparison to the unpaired state.

This conclusion has a clear physical meaning. Indeed, the spin order of unpaired electrons breaks due to thermal fluctuations, forming in real space either a singlet or a triplet. Therefore, there is a probability that the unpaired electrons are a triplet and cannot simultaneously occupy a lowest energy state. One of the unpaired electrons must take a higher state, increasing the total system energy. In contrast to that, paired singlet electrons can simultaneously and permanently occupy a spatial ground state, so the average energy of paired electrons is lower than the energy of unpaired electrons. Then the pairing energy is the difference between two energies: (i) energy of permanent singlet in the spatial ground state; (ii) energy of two unpaired electrons avoiding the same spatial ground state. The negative sign of exchange energy indicates that the singlet has a lower energy than the unpaired state, since the latter contains a non-zero probability of triplet. Thus, the pairing is a direct consequence of the Pauli Exclusion Principle.

If the overlap area of $e_1$, $e_2$ in real space is negligible (i.e. integral $\langle \Psi_1(\vec{r}) | \Psi_2(\vec{r}) \rangle$ is small), then the exchange energy in Eq. (2.4) is negligible. In this case there is no advantage of the singlet state, since the electrons are separated in real space. Thus, the larger the overlap, the greater the energy advantage of the pairing. Consequently two paired wave functions tend to a full coincidence in real space and remain together in equilibrium. Finally two wave packets stay together because their singlet state and full overlap in real space reduce their total energy. Thus, the assumption in Eq. (2.6) is justified.

It is possible to show that the singlet pairing of some conduction electrons is favorable for the whole crystal. We define all parts of the overall energy of two unpaired conduction electrons $\langle E(\vec{r}_1, \vec{r}_2) \rangle$:

1. Kinetic energies of electrons $e_1$ and $e_2$, $\langle K(e_1) \rangle$, $\langle K(e_2) \rangle$;
2. Potential energy of repulsion of electron $e_1$ from all conduction electrons in the crystal, $\langle P(e_1, e) \rangle$;
3. Potential energy of repulsion of electron $e_2$ from all conduction electrons in the crystal, $\langle P(e_2, e) \rangle$;
4. We must correct double counting the repulsion between $e_1$, $e_2$, so we subtract the potential energy of repulsion between electrons $e_1$, $e_2$, $-\langle P(e_1, e_2) \rangle$;
5. Potential energy of attraction of electron $e_1$ to all ions in the crystal, $\langle P(e_1, I) \rangle$;
6. Potential energy of attraction of electron $e_2$ to all ions in the crystal, $\langle P(e_2, I) \rangle$.

The exchange energy of $e_1$, $e_2$ in Eq. (2.9) is a sum of the points 1-6:

$$J = \langle E(\vec{r}_1, \vec{r}_2) \rangle = \langle K(e_1) \rangle + \langle K(e_2) \rangle + \langle P(e_1, e) \rangle + \langle P(e_2, e) \rangle - \langle P(e_1, e_2) \rangle + \langle P(e_1, I) \rangle + \langle P(e_2, I) \rangle \qquad (2.10)$$

The points 1-6 are a list of the crystal energy terms, which contain the paired electrons $e_1$, $e_2$. If the crystal has many singlet pairs, then the energy of each pair contains the points 1-6 (however, we must again correct double counting the repulsion between electrons of different pairs).

The total energy of the many-body crystal contains additional energy terms:

7. Kinetic energies of single conduction electrons;
8. Potential energy of repulsion between single conduction electrons;
9. Potential energy of attraction of single conduction electrons to ions;
10. Potential energy of repulsion between ions.

The points 1-10 are a full list of all crystal energy terms.

The single electrons don't change their states, since they remain unpaired; hence the crystal energy terms in the points 7-10 remain unchanged. In the points 1-6 the overall energy of every singlet pair ($e_j$, $e_i$) is lower than the energy of two unpaired electrons $e_j$, $e_i$ due to the negative exchange energy $J$. Thus, the singlet pairing of some conduction electrons inevitably leads to the energy lowering of the whole crystal, the macroscopic state can exist.

The same crystal energy lowering can be derived by exploring the crystal many-body wave function as a product of accurate eigenfunctions [9], [10] of single and paired electrons (we assume the eigenfunctions are known):

$$\langle E_{crystal} \rangle = \langle ...\varphi_{i-1}(\vec{r}_{i-1}) \cdot \varphi_i(\vec{r}_i) ... \Psi_{m-1}(\vec{r}_{m-1}) \cdot \Psi_m(\vec{r}_m) ... | H(\vec{r}_1 ... \vec{r}_n) | ... \varphi_{i-1}(\vec{r}_{i-1}) \cdot \varphi_i(\vec{r}_i) ... \Psi_{m-1}(\vec{r}_{m-1}) \cdot \Psi_m(\vec{r}_m) ... \rangle \qquad (2.11)$$

Where: $\varphi_{i-1}(\vec{r}_{i-1}), \varphi_i(\vec{r}_i)$ accurate normalized eigenfunctions of **paired** electrons; $\Psi_{m-1}(\vec{r}_{m-1}), \Psi_m(\vec{r}_m)$ accurate normalized eigenfunctions of **unpaired** electrons; $\vec{r}_1 ... \vec{r}_n$ radius-vectors of electrons and ions; $H(\vec{r}_1 ... \vec{r}_n)$ crystal Hamiltonian containing operators for energies:

1. Kinetic energies of paired and single conduction electrons;
2. Potential energy of repulsion between all conduction electrons (paired and single);
3. Potential energy of attraction to ions of all conduction electrons (paired and single);
4. Potential energy of repulsion between ions.

All conclusions from Eqs (2.1)-(2.10) are valid if electrons $e_1$, $e_2$ are two equal running Bloch waves [11]:

$$\Psi_1(\vec{r}, t) = \Psi_2(\vec{r}, t) = u(\vec{r}) \cdot \exp(-i\omega t - i\vec{k}\,\vec{r}) \qquad (2.12)$$

However, it is a rare event that the momenta $\vec{k}$ of two running waves are equal permanently in time.

The momenta of electrons may be permanently equal if before pairing each electron is a **standing wave**, which is a sum of two equiprobable Bloch waves propagating in opposite directions:

$$\Psi_1(\vec{r},t)=\Psi_2(\vec{r},t)=\frac{1}{\sqrt{2}}u(\vec{r})\cdot\exp(-i\omega t-i\vec{k}\,\vec{r})+\frac{1}{\sqrt{2}}u(\vec{r})\cdot\exp(-i\omega t+i\vec{k}\,\vec{r}) \qquad (2.13)$$

The total momentum of each standing wave is zero, hence the momenta of electrons are synchronous, so the pairing is possible despite the fact that the kinetic energy of electrons may be larger than their pairing energy.

The overall energy of two unpaired electrons $\langle E(\vec{r}_1,\vec{r}_2)\rangle$ is usually not arbitrarily small; consequently the exchange energy $J$ in Eq. (2.9) is also not arbitrarily small. The sign of the exchange term $J$ is related with the sign of the energy increment resulting from the pairing. This energy increment is related with wave function modifications and is not necessarily negligible if $J$ is not negligible; therefore, the binding energy 2Δ in the singlet pair is also not negligible:

$$2\Delta=\left\langle\varphi_1(\vec{r}_1)\varphi_2(\vec{r}_2)\middle|\hat{O}(\vec{r}_1,\vec{r}_2)\middle|\varphi_1(\vec{r}_1)\varphi_2(\vec{r}_2)\right\rangle-\left\langle\Psi_1(\vec{r}_1)\Psi_2(\vec{r}_2)\middle|\hat{O}(\vec{r}_1,\vec{r}_2)\middle|\Psi_1(\vec{r}_1)\Psi_2(\vec{r}_2)\right\rangle=A\cdot J \qquad (2.14)$$

Where $\varphi_1(\vec{r}_1),\varphi_2(\vec{r}_2)$ are eigenfunctions of paired $e_1$, $e_2$; $A$ is a positive material constant.

Note, Δ is the energy for transfer of one electron from its ground state (paired $\varphi$ state) into the nearest neighboring excited state (unpaired Ψ state). If the ground state is local, then the excitation Δ is not necessarily negligible.

The binding (pairing) energy 2Δ is not arbitrarily weak, so the unpaired states of $e_1$, $e_2$ are **instable** as excited states. However, the paired state is lasting only if external energies (temperature, radiation) are weaker than the excitation energy Δ. The excitation energy Δ is roughly related to the **pairing temperature** $T^*$; $|\Delta|\approx k_BT^*$.

If the wave functions of two conduction electrons in crystal (for sample two s-electrons) permanently coincide in real space and form a permanent singlet, then the electrons are similar to the electrons in the ground state of Helium. The difference is that in crystal the wave functions cover many ions and the pair can tunnel along crystal, since all crystal areas are equipotential for the pair. In the ground state of Helium the singlet state is favorable despite the fact that the repulsion of electrons is maximal; the increase in attraction of the singlet s-electrons to the Helium nucleus exceeds the increase in repulsion and in kinetic energy.

The electron pair is stable like a valence bond in multi-atom systems, so the pair doesn't form/lose any bonds in the crystal and doesn't absorb/radiate any chemical energy.

Every standing wave is limited in real space, therefore, a stable singlet of two standing waves can be considered as a zero-spin **boson**. Bosons can form the Bose-Einstein-Condensate (BEC) below a certain temperature $T_{\mathrm{BEC}}$ [12]. If all electron pairs are in the BEC ground state, then the excitation energy of every pair is related to $k_{\mathrm{B}}T_{\mathrm{BEC}}$, which is not zero if the bosonic density is not zero. If all external influences are weaker than $k_{\mathrm{B}}T_{\mathrm{BEC}}$, then the pairs remain in the BEC ground state and cannot absorb/radiate any energy in the crystal; as a result the total energy and momentum of all pairs don't dissipate, the pairs fluctuate without resistance. Thus, the pairing of conduction electrons and BEC of the pairs lead to the zero resistivity (likewise works the superfluidity in Helium-4; the pairing energy of electrons $k_BT^*$ in Helium-4 is huge, whereas $k_{\mathrm{B}}T_{\mathrm{BEC}}$ is small).

In an external magnetic field $H$ the crystal obtains an additional energy density $w=0{,}5\cdot\mu_0\mu H^2$. The non-dissipative pairs fluctuate in the field $H$ as free particles with a charge $-2e$ and zero spin. Consequently the non-dissipative fluctuations of pairs can be redistributed into non-dissipative currents suppressing the additional magnetic energy $w$ (Meissner effect).

If the binding energy |Δ| per one electron in Eq. (2.14) is larger than the insulating band gap $E_g$ of the crystal, then electrons can leave the valence band at a temperature $T<T^*$, hence the electrons may pair up despite the band gap. A doping in the crystal may reduce the band gap and, thus, give rise to the pairing. This doping effect is observable in cuprates [13], in iron-based superconductors [14], in semiconductors [15].

We can show a necessary condition for SC in metals is that the electrons, before pairing, are close to the Fermi surface in the momentum space. Imagine that the pairing occurs when the energy of single electrons has a value $E_\lambda$ much lower than the Fermi energy $E_F$. If the thermal energy doesn't exceed the binding energy $|\Delta|$ in Eq. (2.14), then the density of pairs is not zero and in the energy spectrum of single electrons occurs a gap around the level $E_\lambda$. The gap is thin, since $|\Delta|$ is usually small and the density of electrons with close energies is limited by the Exclusion Principle. $E_\lambda$ is notably less than $E_F$, so there are single electrons with energies larger than $E_\lambda$. These single electrons may drop to the pairing level $E_\lambda$ due to energy fluctuations and may, thus, form new pairs. The density of paired electrons is limited by the thin gap around $E_\lambda$; therefore, the new pairs replace the already existing pairs, and the latter lose their paired state. Thus, each electron is **not permanently** paired, but it becomes periodically unpaired. In the unpaired state the electron is normal (resistive), therefore, the momentum of the electron and of the pair dissipates. Thus, the case $E_\lambda < E_F$ cannot keep a permanent supercurrent, although the pairing is possible. If $E_\lambda = E_F$ then every pair may exist **permanently in time**, because below a temperature $T_c$ all single electrons are below $E_\lambda$ and cannot overcome the energy gap and cannot reach the pairing level $E_\lambda = E_F$; as a result new pairs don't arise and don't replace existing pairs. Hence the pair recombination doesn't occur and the total momentum of all pairs doesn't dissipate. Thus, the superconducting pairing happens only for single electrons in an energy gap with $E_F$ **as the upper limit**. Only such permanent pairs are superconducting. At temperature above $T_c$ the thermal energy is sufficient to transfer single electrons through the energy gap to the pairing level $E_\lambda$, therefore, new pairs arise and replace the existing ones, the state becomes dissipative. We note, the permanently paired electrons and single electrons are **distinguishable (non-interchangeable) in the momentum space**; the same particle distinguishability is valid in systems containing free electrons and electrons bound in atoms/molecules below their ionization temperature.

3. **Pairing of standing waves.**

We found that the binding energy in a singlet pair $e_1$, $e_2$ is maximal if the overlap integral $\langle \Psi_1(\vec{r}) | \Psi_2(\vec{r}) \rangle$ is maximal, i.e. the wave functions coincide in real space. The electron motion can split the pair in real space, because the energy gap of superconductors has order of magnitude $10^{-3}$ eV, while $E_F$ is usually much larger (a few eV). The energy of very slow electrons is usually much lower than the Fermi energy; hence the slow electrons cannot form superconducting pairs. Two electrons can form a pair if their momenta are synchronous, but it is a rare event for running waves. The electrons as **standing waves** have zero-momenta, hence their momenta may be synchronous and the pairing is possible despite a large kinetic energy.

Standing waves arise as a result of multiple reflections from a spatial potential, so a standing wave is limited in real space, i.e. the standing state is a **local** potential well. The excitation and delocalization energy of a local electron is not necessarily negligible, hence the pair-breaking energy is also not negligible. Note, the pairing of electrons as standing waves is a local event, while electron pairs as condensed bosons of a Bose-Einstein-Condensate are no longer local, since the BEC is a macroscopic coherence of all pairs.

The condition of the standing wave in a crystal is the Bragg condition [16]:

$$n\lambda = 2 \cdot R \qquad (3.1)$$

Where: $n$ integer; $\lambda$ length of the Bloch wave in Eq. (2.12); $R$ one of lattice parameters.

Under Bragg condition the electron becomes a set of standing waves with a zero total momentum [17].

At $n=1$ in Eq. (3.1) the length of the standing wave is maximal: $\lambda_1=2R$. Each crystal has some values $R$; ($R_{100}$, $R_{110}$, $R_{210}$ etc. depending on crystal axis) and, thus, some values $\lambda_1$. Each value $\lambda_1=2R$ is linked to the energy $E_{\lambda=2R}$ :

$$E_{\lambda=2R} = \frac{(h/\lambda_1)^2}{2 \cdot m} = \frac{h^2}{R^2 \cdot 8 \cdot m} \qquad (3.2)$$

Where $m$ is the inertial mass of free electron.

Not all materials have conduction electrons with short values $\lambda=2R$ and with its associated $E_{\lambda=2R}$ . If the Fermi energy of a crystal is low, then $\lambda$ values are larger than short values $2R$ and electron energies are lower than corresponding $E_{\lambda=2R}$ ; so the states with short values $\lambda=2R$ are empty and short standing waves don't occur. In some metals $E_{\lambda=2R}$ is close to $E_F$ (it is equivalent that $\lambda_F=2R$). Probably in some crystals the formation of pairs is possible at $n$ larger than 1 in Eq. (3.1). For sample at $n=2$ the length of standing waves is $\lambda_2=R$.

Every act of the pairing is energetically favorable; hence the energy $2|\Delta|$ is emitted. Thus, the momentum of each pair increases from zero to a value $p \le (2|\Delta| \cdot 2 \cdot 2m)^{0.5}$. This momentum $p$ should be added to the momenta of paired electrons in the momentum space. Therefore, kinetic energies of electrons from different pairs are not equal, but distributed into a spectrum with the width $\approx |\Delta|$. For instance, if the kinetic energy of normal electrons was $E_F$ before pairing, then after the pairing the kinetic energy becomes $\approx E_F + |\Delta|$. So the spectrum of paired electrons is **above** the spectrum of single electrons in the momentum space. Note: the same conclusion follows from the exchange energy; the exchange term for kinetic energy is positive, i.e. the kinetic energy grows by pairing, while the total energy falls (the same relation is valid for all singlet bonds in chemistry). Thus, the paired electrons can overlap with normal electrons in real space. The density of paired electrons is $\approx S(E_F) \cdot |\Delta|$, where $S(E_F)$ is the density of states of electrons on Fermi surface. The spectrum of normal electrons obtains a corresponding gap $(E_2 - E_1) \approx |\Delta|$ around the value $E_{\lambda=2R}$ ($E_1$, $E_2$ are limits of the gap). The gap is not negligible if the thermal energy is insufficient to destroy the pairs. As shown above a necessary condition for SC is that $E_F$ is the upper limit of the gap: $E_F = E_2$.

The energy gap is $(E_F - E_1)$ , where the gap bottom $E_1$ should be below $E_{\lambda=2R}$ (otherwise new pairs arise and replace the existing ones, energy dissipates). The density of the singlet electrons $N_s$ is limited by the energy gap:

$$N_s = \int_{E_1}^{E_F} S(E_\lambda)\, dE_\lambda \qquad (3.3)$$

We note that the formation of singlet pairs in a thin layer on the Fermi surface does not affect the wave function symmetry of normal electrons below the gap $(E_F - E_1)$, so the states and energies of normal electrons are unaltered, as assumed in Eq. (2.11).

The electrons (before superconducting pairing) must be close to the Fermi surface, i.e. the value $E_{\lambda=2R}$ must be close to $E_F$ (i.e. $\lambda_F \approx 2R$). Indeed, the energy gap is much less than $E_F$ ; therefore, if $E_{\lambda=2R}$ is significantly less than $E_F$ , then the upper gap limit $E_2$ is also less than $E_F$ ; as shown above this case is not superconducting because the pairs are not permanent in time. For this reason Au, Ag, Cu (where $E_{\lambda=2R} < E_F$ significantly) are not superconductors. If $E_{\lambda=2R}$ is significantly larger than $E_F$ , then there are no electrons with $\lambda$=2$R$ and the gap doesn't occur. For this reason in some structures with a low $E_F$ a doping may raise the carrier density and its associated $E_F$ up to the level $E_{\lambda=2R}$ (which is constant, if $R$ doesn't change). Thus, the doping may lead to SC, $T_c$ increases. If the crystal is overdoped, then $E_F$ is too large; $E_{\lambda=2R} < E_F$ , $T_c$ vanishes. This doping effect explains the $\boldsymbol{T_c}$ **- dome** of phase diagrams of superconductors [18]. Thus, a large value $|E_F - E_{\lambda=2R}|$ suppresses $T_c$. If $E_F = E_{\lambda=2R}$, then $T_c$ corresponds to the pairing energy $\Delta$ in Eq. (2.14); however, the doping influences on both $E_F$ and $\Delta$, so the $T_c$ - maximum is not always pinned exactly to $E_F = E_{\lambda=2R}$ . A double dome is possible due to the fact that the crystal has some lattice parameters.

We can specify the energy $Ck_BT_c$ ($C$ is a material specific constant) as thermal energy, which is necessary to scatter single electrons from the SC-gap bottom to the pairing level $E_{\lambda=2R}$ , where new pairs arise and replace the old ones. Thus, $Ck_BT_c$ is an energy area between the SC-gap bottom and $E_{\lambda=2R}$, i.e. $Ck_BT_c$=($E_{\lambda=2R}$ - $E_1$). So we know about the $T_c$ tuning:

$$T_c = 0 \qquad if \;\; |\Delta| < |E_F - E_{\lambda=2R}| \qquad (3.4)$$

$$C\,k_BT_c \le |\Delta| \qquad if \;\; |\Delta| > |E_F - E_{\lambda=2R}| \;\; and \;\; E_{\lambda=2R} \ge E_F \qquad (3.5)$$

$$C\,k_BT_c \le |\Delta| - |E_F - E_{\lambda=2R}| \qquad if \;\; |\Delta| > |E_F - E_{\lambda=2R}| \;\; and \;\; E_{\lambda=2R} < E_F \qquad (3.6)$$

A growing $T$ continuously fills up and closes the area $Ck_BT_c$ by putting there single (normal) electrons. At $T > T_c$ the single electrons can reach the level $E_{\lambda=2R}$. Knowing the density of single states $S(E_\lambda)$ around the SC-gap, we can calculate $T_c$ and $C$ - parameter in $Ck_BT_c$.

The pairing energy $\Delta$ can be found from Eq. (2.14) as the difference between the average energy of ground (paired) states and nearest neighboring excited (unpaired) states. The SC - gap bottom $E_1$ is to find from Eqs (3.5) or (3.6):

$$E_1 = E_{\lambda=2R} - C\,k_BT_c = E_{\lambda=2R} - |\Delta| \qquad if \;\; E_{\lambda=2R} \ge E_F \qquad (3.7)$$

$$E_1 = E_{\lambda=2R} - C\,k_BT_c = E_F - |\Delta| \qquad if \;\; E_{\lambda=2R} < E_F \qquad (3.8)$$

The single-electron-distribution on the SC-gap bottom $E_1$ is the Fermi-Dirac function $f(E_\lambda,\ T)$. The single-electron-density $N(T_c)$ at $T$ just below $T_c$ is:

$$N(T_c) = \int_0^{E_{\lambda=2R}} S(E_\lambda)\cdot f(E_\lambda, T_c)\,dE_\lambda = \int_0^{E_{\lambda=2R}} S(E_\lambda)\cdot \frac{1}{\exp\left(\dfrac{E_\lambda - E_1}{k_BT_c}\right)+1}\,dE_\lambda \qquad (3.9)$$

At $T=0$ the single-electron- density $N$(0) is:

$$N(0) = \int_0^{E_1} S(E_\lambda)\,dE_\lambda \qquad (3.10)$$

New pairs don't arise at $T < T_c$, hence $N(0)=N(T_c)$ just below $T_c$ and Eq. (3.9) is equal to Eq. (3.10):

$$\int_0^{E_1} S(E_\lambda)\,dE_\lambda = \int_0^{E_{\lambda=2R}} S(E_\lambda)\cdot \frac{1}{\exp\left(\dfrac{E_\lambda - E_1}{k_BT_c}\right)+1}\,dE_\lambda \qquad (3.11)$$

$E_{\lambda=2R}$ and $E_F$ are known from crystal structure, so $E_1$ is calculable from Eqs (3.7) or (3.8) with values $E_{\lambda=2R}$, $E_F$, $\Delta$. Hence, knowing $S(E_\lambda)$, we can calculate $T_c$ from Eq. (3.11).

$C$ parameter in $Ck_BT_c$ we find from $T_c$ and using $Ck_BT_c$=($E_{\lambda=2R}$ - $E_1$). Calculations with Eq. (3.11) show:

**A**. If $S(E_\lambda)$ is proportional to $E_\lambda^{0,5}$ (as in Fermi liquids), then $C$ slightly depends on values $E_1$ and $T_c$. Substituting the ranges $E_1$=(0.25 - 3) eV and $T_c$=(0.1 - 50) K into Eq. (3.11) we find $E_{\lambda=2R}$ and the range $C \approx$ (4.5 - 11) units;

**B**. $C$ depends on the $S(E_\lambda)$-slope around the level $E_1$: the larger $dS/dE_\lambda$, the smaller $C$. On the zone edge ($\lambda=2R$) $dS/dE_\lambda$ may be larger than the $E_\lambda^{0,5}$ -slope, therefore, $C$ may be smaller than (4.5 - 11) by a few units, i.e. $C \approx$ (3 - 7). These $C$ - values are consistent with experiments.

The isotope substitution is a way to tune $T_c$ by tuning $E_F$ to $E_{\lambda=2R}$ based on the fact that $E_F$ depends on the effective mass of electron $m^*$ and electron density $N$ [19], whereas $E_{\lambda=2R}$ in Eq. (3.2) depends only on the lattice parameter $R$.

$$E_F = \frac{h^2}{8m^*}\left(\frac{3N}{\pi}\right)^{2/3} \qquad (3.12)$$

The isotope effect is a consequence that the energy of phonons is proportional to $M^{-0,5}$ ($M$ - mass of ion). The decrease in $M$ raises the energy of phonons; therefore, the electron-ion interaction and its associated reflection of electrons from ions may intensify. This reflection intensification is equivalent to the increase in the effective mass $m^*$ and, thus, to the decrease in $E_F$, whereas $R$ is almost unchanged. If the initial value $E_F$ is larger than $E_{\lambda=2R}$ (it is usual for metals), then the decrease in $M$ pulls $E_F$ down closer to $E_{\lambda=2R}$; hence $T_c$ grows (the isotope coefficient α>0). If the initial value $E_F$ is less than $E_{\lambda=2R}$, then the decrease in $M$ pulls $E_F$ down away from $E_{\lambda=2R}$; hence $T_c$ may vanish (α<0). One can conclude that in case $E_F \approx E_{\lambda=2R}$ the isotope effect may be weak ( $|\alpha|$ <0,5). Thus, the different values and sign of α [20] are a result of the different initial positions $E_F$ to $E_{\lambda=2R}$.

Other ways to tune $T_c$ by tuning $E_F$ to $E_{\lambda=2R}$ are: electric field [21] since $E_F$ depends on electron density; film thickness [22], [23], [24], [25] since the mutual $E_F$ of layered heterostructures depends on layer thicknesses; high pressure [26], [27], [28] since $E_F$ depends on the distance between atoms.

A further sample of $E_F$ tuning is the alkali metals Li, Na, K, Rb, Cs (table 1). Only Li is superconductor at ambient pressure [29] and only Li has $E_{\lambda=2R}$ =3,09 eV (calculated by Eq. (3.2) in bcc-structure, $R_{100}$=3,49 Å) relatively close to $E_F$ ≈3,2 eV [30] at ambient temperature. The next candidate in superconductors is Cs: $E_{\lambda=2R}$ =1,33 eV (calculated by Eq. (3.2) in bcc-structure, $R_{111}=6{,}14\cdot 3^{1/2}/2=5.32$ Å), $E_F$ ≈1,54 eV calculated by Eq. (3.12); Cs is really superconductor at 12 GPa [31]. The trend is confirmed by Rb; the gap between $E_{\lambda=2R}$ and $E_F$ is larger than in Cs, so SC occurs at a larger pressure, 55 GPa [32]. The other alkali metals are not superconductors and their values $E_F$ are larger than $E_{\lambda=2R}$ more significantly than in Li, Cs, Rb. The high pressure increases the density of ions, so $m^*$ rises and $E_F$ drops to $E_{\lambda=2R}$ ; therefore, $T_c$ grows in Li, Cs, Rb. We note that $E_F$ and $E_{\lambda=2R}$ both are proportional to $R^{-2}$, hence without modification of $m^*$ an isotropic $R$-reduction increases both $E_F$ and $E_{\lambda=2R}$ .

**Table 1**. *Comparison of energies $E_F$ and $E_{\lambda=2R}$ for alkali metals. $E_{\lambda=2R}$ are calculated by Eq. (3.2) for lattice parameters $R_{100}$ and $R_{111}$ in bcc crystals. Larger R-values are not considered, since corresponding $E_{\lambda=2R}$ values are much lower than $E_F$. For Li is used the experimental value $E_F$ at ambient temperature; for other alkali metals are used $E_F$ values calculated by Eq. (3.12) corresponding roughly to experimental values.*

| | Discovered SC | $E_F$ | | $E_{\lambda=2R}$ for $R_{100}$ | $E_F$ - $E_{\lambda=2R}$ for $R_{100}$ | | $E_{\lambda=2R}$ for $R_{111}$ | $E_F$ - $E_{\lambda=2R}$ for $R_{111}$ |
|---|---|---|---|---|---|---|---|---|
| | | eV | | eV | eV | | eV | eV |
| **Li** | SC | 3.2 | | 3.09 | **0.11** | | 4.12 | -0.92 |
| **Na** | | 3.16 | | 2.05 | 1.11 | | 2.73 | 0.43 |
| **K** | | 2.04 | | 1.32 | 0.72 | | 1.76 | 0.28 |
| **Rb** | SC at 55 GPa | 1.78 | | 1.15 | 0.63 | | 1.54 | **0.24** |
| **Cs** | SC at 12 GPa | 1.54 | | 1.0 | 0.54 | | 1.33 | **0.21** |

The approach explains the combined isotope and high pressure effect in lithium [33]. In $^6$Li the high pressure and light isotope pull $E_F$ below the level $E_{\lambda=2R}$, so $T_c$ starts to diminish at a certain pressure $p_0$. In heavy lithium-7 $E_F$ at $p_0$ remains above $E_{\lambda=2R}$, hence the increasing pressure continues to pull $E_F$ down closer to $E_{\lambda=2R}$; $T_c$ continues to grow. As a result the sign of $dT_c/dp$ above the pressure $p_0$ is different for $^6$Li and $^7$Li.

A perfect conductor cannot form the Cooper pairs, because the electrons pass through the lattice without reflection, standing local states don't arise, electronic wave packets are unlimited in real space, hence a correlation of 2 electrons is weak and negligible thermal fluctuations prevent the pairing. So in Eq. (2.14) the factors $A$ and $J$ vanish; the pairing energy $2\Delta=A\cdot J$ also vanishes. Thus, the pairing energy should be related with the strength of the electron reflection via potential energy of electrons in Eq. (2.10). A deeper potential energy leads to a deeper $J$ in Eq. (2.10) and, thus, to a stronger pairing energy in Eq. (2.14). On the other hand, a deeper average potential energy means a deeper potential on each ion, which reflects conduction electrons more strongly, so the electronic wave packets become shorter. Factor $A$ in Eq. (2.14) is maximum for short range wave packets (for instance, in $H_2$ molecule), so the shorter the wave packets, the larger the factor $A$. Thus, the singlet bond and $T_c$ may be stronger if the electron reflection is stronger, but under the condition $E_F = E_{\lambda=2R}$. Note, lattice vibrations may boost the electron reflection, enhancing the pairing strength.

The approach is consistent with the fact that high temperature superconductors are layered structures and poor conductors in the normal state. In layered structures is possible to combine two poorly compatible things: a large effective mass $m^*$ (related to the strong electron-ion interaction/reflection) and a large $E_F$ (up to the value $E_{\lambda=2R}$). This is because $E_F$ in thin films is larger than in bulk [34], whereas the electron reflection and $m^*$ in-plane may remain almost unchanged. In a 3-dimensional structure is difficult to combine a large $m^*$ (> $5\cdot m$) and $E_F \approx E_{\lambda=2R}$ (a few eV). Thus, $T_c$ in quasi 2-dimensional systems can be higher.

The pairing energy in the proposed model is related rather to the lattice potential than to the carrier density. This enables SC at relatively low carrier densities. A sample is superconducting bismuth at ambient pressure, a semimetal with a low carrier density, $N \approx 3\cdot10^{17}$ /cm$^3$ [35]. $E_F$ of bismuth is ≈ 25 meV, hence corresponding $\lambda_F = h\cdot(E_F 2m)^{-0.5} \approx 0.78\cdot10^{-8}$ m. Thus, in bismuth work the long values of $R=0.5\cdot\lambda_F \approx 0.39\cdot10^{-8}$ m (i.e. $R_{810}$, $R_{910}$ and similar). These long standing waves exist only due to the high crystal purity, where the electronic mean free path is much larger than $\lambda_F$. Thus, SC-pairs emerge from the long standing waves on the Fermi surface. Since the electron's waves are long range, the mean distance between electrons in one pair may be larger than the electron's wave length $\lambda_F \approx 10^{-8}$ m, so the pairing is possible at a carrier density less than $(\lambda_F)^{-3} \approx 10^{18}$ /cm$^3$.

For Fermi metals we can estimate the relation between BEC-temperature and pairing temperature, $T_{BEC}/T^*$. Assuming $N_s \approx S(E_F)\cdot|\Delta|$ and $3.5\cdot k_B T^* \approx |\Delta|$, using the well-known equations $S(E_F) = 2^{1/2} m^{3/2} E_F^{1/2} / (\pi^2 \hbar^3)$ for Fermi liquid and $T_{BEC}(N_s) \approx 3.3125\cdot \hbar^2 (N_s/2)^{2/3} / (2m k_B)$ for boson gas we obtain:

$$\frac{T_{BEC}}{T^*} \approx \left(\frac{E_F}{|\Delta|}\right)^{1/3} \qquad (3.13)$$

For Fermi metals usually $E_F >> |\Delta|$, hence $T_{BEC} > T^*$ and SC depends rather on $T^*$ than on $T_{BEC}$, i.e. $T_c = T^* < T_{BEC}$.

Eq. (3.13) is not appropriate for strongly correlated systems, because $S(E_F)$ is usually smaller than $S(E_F)$ of Fermi liquid and $E_F$ may be order of magnitude $|\Delta|$, so BEC may define SC, i.e. $T_c = T_{BEC} < T^*$. Consider $E_F$ is close to the SC-gap bottom $E_1$ and notably below $E_{\lambda=2R}$ (that is possible, for example, in underdoped cuprates). Then in Eq. (3.3) the gap ($E_F$ - $E_1$) and the corresponding singlet density $N_s$ are small. Hence the resultant $T_{BEC}$ may be lower than the pairing temperature $T^*$, because $T^*$ depends rather on the lattice potential than on $N_s$. If the BEC is a necessary condition for SC and $T_{BEC} < T^*$, then $T_c = T_{BEC}$; so $T_c$ is also lower than $T^*$. Thus, at $T$ between $T_c$ and $T^*$ there are permanent pairs without SC. This may be related with the pseudogap in some superconductors, as observed in [36].

If the pairing temperature $T^*$ is larger than $T_{BEC}$ (i.e. $T^* > T_{BEC}$), then just above $T_{BEC}$ the electron pairs may be permanent but non-superconducting. The non-superconducting pairs are observed in [37], [38].

The room temperature SC in Ag-Au nanostructures [39], [40] points to the local pairing and BEC of the pairs. Indeed, local states can occur within Ag cluster, because the Volta potential in Ag is lower than in Au by roughly 0.7 eV, so a multiple electron reflection occurs on the Ag/Au interface, localizing electrons. A permanent singlet occurs within an Ag cluster, when the pairing energy $k_B T^*$ is larger than thermal fluctuations $k_B T$. The density of Ag-nanoparticles corresponds to a boson density, which is necessary for $T_{BEC}$ = 286 K, as calculated in [41]. Moreover, $T_c$ grows with increasing Ag-nanoparticle density. Hence one can conclude that $T_c = T_{BEC} < T^*$, the local pairing is pinned to each Ag-nanoparticle, $T_{BEC}$ and, thus, $T_c$ both are related to the Ag-nanoparticle density.

## 4. Conclusion and discussion.

The above argumentation shows that the exchange interaction may cause the electron pairing in crystals. The exchange energy of local electrons is not necessarily negligible; hence the pairing energy is also not negligible. Thus, the non-zero pairing energy is a result of the Pauli Exclusion Principle.

Further necessary conditions for SC: stability of every electron pair (non-breaking pairs provided with pairing of standing waves on the Fermi surface); non-zero temperature of the BEC ground state of the electron pairs.

The approach of the exchange energy is clearly applicable when the waves $\Psi_1(\vec{r}_1)$, $\Psi_2(\vec{r}_2)$ contain s-orbitals, because the s-orbitals envelop each ion and the singlet pairing leads to a convergence of electrons to ions. In case of p-, d-, f-orbitals the described approach works if the orbitals envelop **nearest neighbor** ions. In this case the singlet pairing depends on the orbital orientation and on factors influencing the distance between ions (pressure, doping etc.)

The approach of standing waves is related with the Bragg-interference, which may form diffraction patterns in crystals. This explains why the charge density order precedes the superconductivity in cuprates [42], [43], [44]. The insulator-superconductor transitions may also be related with the local standing waves preceding the electron pairing.

Following the proposed approach we can define main ways to a higher $T_c$ in new superconductors:

A. The value $E_F$ should be tunable close to highest $E_{\lambda=2R}$ (i.e. $\lambda_F$ is tunable to shortest values $2R$). The tuning is possible by doping, pressure, film thickness, electric field etc.
B. The material should have a large value $m^*$, since $m^*$ is related to deeper lattice potentials in Eq. (2.10) causing a deeper exchange energy. However, the condition $E_F = E_{\lambda=2R}$ should be kept. Low dimensionality can help to combine large $E_F$ and m* values. Superlattices and host lattices can reduce $E_{\lambda=2R}$ to low $E_F$ values.
C. The material should be microscopically homogeneous, because the disorder scatters the electron localization length, so the pairing energy, pair density and related $T_c$ become inhomogeneous in real space, reducing the maximum $T_c$.

## 5. References.